\documentclass[aps,prd,11pt,preprint,nofootinbib,showkeys]{revtex4-1}
\usepackage{amsmath,amssymb,amsthm,mathrsfs,bbm}
\usepackage{latexsym,amscd,amsbsy,amsfonts,dsfont}
\usepackage{graphicx}
\usepackage[utf8]{inputenc}
\usepackage{subfigure}  
\usepackage{float}
\usepackage{slashed}
\usepackage{cancel}
\usepackage{multirow}
\usepackage{soul}
\usepackage{physics}
\usepackage{comment}
\usepackage[usenames,dvipsnames]{xcolor}
\usepackage[
      colorlinks=true,
      linktocpage=true,
      breaklinks=true,
      linkcolor=Aquamarine,
      urlcolor=Purple,
      filecolor=black,
      citecolor=Purple,
      menucolor=black,
      pdfstartview=FitV,
      bookmarksopen=true
      ]{hyperref}

\makeindex

\begin{document}

\title{Embedding Unimodular Gravity in String Theory}

\author{Luis J. Garay}
\email{luisj.garay@ucm.es}
\affiliation{Departamento de F\'{\i}sica Te\'orica and IPARCOS, Universidad Complutense de Madrid, 28040 Madrid, Spain}
\author{Gerardo Garc\'ia-Moreno}
\email{ggarcia@iaa.es}
\affiliation{Instituto de Astrof\'{\i}sica de Andaluc\'{\i}a (IAA-CSIC), Glorieta de la Astronom\'{\i}a, 18008 Granada, Spain}

\preprint{IPARCOS-UCM-23-001}
\begin{abstract}
Unimodular Gravity is a theory displaying Weyl rescalings of the metric and transverse (volume-preserving) diffeomorphisms as gauge symmetries, as opposed to the full set of diffeomorphisms displayed by General Relativity. Recently, we presented a systematic comparison of both theories, concluding that both of them are equivalent in everything but the behaviour of the cosmological constant under radiative corrections. A careful study of how Unimodular Gravity can be embedded in the string theory framework has not been provided yet and was not analyzed there in detail. In this article, we provide such an explicit analysis, filling the gap in the literature. We restrict ourselves to the unoriented bosonic string theory in critical dimension for the sake of simplicity, although we argue that no differences are expected for other string theories. Our conclusions are that both a Diff and a WTDiff invariance principle are equally valid for describing the massless excitations of the string spectrum. 
\end{abstract}


\maketitle
 
\tableofcontents

\section{Introduction}
\label{Sec:Introduction}

Unimodular gravity (UG) is a theory which is so similar to General Relativity (GR) that one may wonder to what extent both of them are equivalent. Recently we presented a systematic comparison of both theories in all the regimes and situations in which a potential difference might appear, which was still lacking~\cite{Carballo-Rubio2022}. We concluded that for all of the possible regimes analyzed there, both theories are equivalent except for the behaviour of the cosmological constant. Whereas the cosmological constant is radiatively stable in UG~\cite{Carballo-Rubio2015} (it is simply an integration constant of the equations of motion), in GR it is radiatively unstable. In this way, if one uses technical naturalness in the sense introduced by 't Hooft~\cite{tHooft1979,Burgess2020,Carballo-Rubio2022} as a guiding principle toward building theories, UG theories are much more desirable than GR theories since the cosmological constant is technically natural. 

There are mainly three arguments used to argue that the low-energy limit of string theory is given by the effective field theory (EFT) consisting of GR coupled to some other fields. First of all, when one analyzes the massless spectrum (leaving aside the tachyon field) of bosonic string theory propagating on top of flat spacetime one finds that for oriented strings it contains a graviton, a Kalb-Ramond field, and a dilaton; and for unoriented strings it contains only a graviton and a dilaton. In principle, for computing observables only involving massless states, one expects that one can write down an effective action which simply involves fields that account for these massless excitations, i.e., a graviton-field $h^{\mu \nu}$, (possibly) a Kalb-Ramond field $B_{\mu \nu}$, and a dilaton field $\Phi$. As usual, the fundamental observable considered is the $S$-matrix. 

Now, we come to the arguments used to argue that GR ``emerges naturally" as the low-energy description of such degrees of freedom. First of all, it has been argued that the only self-consistent way of coupling the graviton (massless spin-2 representation of the Poincaré group) to itself is through GR. In that way, having a massless spin-2 field in the spectrum, one necessarily guesses that the non-linear structure of the theory needs to be GR up to potential higher-derivative corrections arising in the EFT. However, we argued~\cite{Carballo-Rubio2022,Delhom2022} that the self-coupling of UG gravitons (those displaying linerarized WTDiff gauge-invariance) to themselves also gives rise to the full non-linear UG in a consistent way, although the coupling of the graviton to itself is through the traceless part of the energy-momentum tensor, instead of the full one. Hence, this first argument does not allow one to discern whether UG or GR is preferred from the string point of view since one is as legitimate as the other. 

The second argument comes from the analysis of string scattering amplitudes, which was already revisited in~\cite{Carballo-Rubio2022}. One can compute within string perturbation theory the scattering amplitudes for graviton asymptotic states. The result is that, to the lowest order in $\alpha'$ and at string tree level, one obtains the same scattering amplitudes obtained in GR. The point is that UG scattering amplitudes are exactly the same as the GR scattering amplitudes~\cite{Alvarez2016,Carballo-Rubio2019}. In that sense, GR is not preferred over UG from the point of view of scattering amplitudes either, as it was concluded in~\cite{Carballo-Rubio2022}.

The final argument comes from analyzing perturbatively in $\alpha'$ the non-linear sigma model that arises from coupling the string degrees of freedom to an arbitrary background metric (or conformal structure), Kalb-Ramond field, and dilaton field generated by the string degrees of freedom themselves. For such a model, the Weyl symmetry of the worldsheet, which is potentially anomalous, needs to be handled carefully. However, although in flat spacetime and zero background fields it simply constrains the dimension of spacetime to be $26$ (critical dimension), in this case constraints also appear for the spacetime fields entering the sigma model construction. Such constraints arise from imposing a cancellation of the Weyl anomaly to make it a sensible theory. The equations that arise are basically Einstein equations, although interpreted as $\beta$-functionals. Both GR and UG give rise to Einstein equations, hence from this point of view we show that it is possible to write both a GR and UG-like EFT for the massless degrees of freedom of the string. Moreover, both actions are also consistent with the previous argument since they reproduce all the scattering amplitudes involving massless states of the string. The only difference that seems to appear, is that, whereas in the GR EFT the cosmological constant is a coupling constant that needs to be set to zero, in the UG EFT it is an integration constant that needs to be set to zero. In other words, UG contains the space of theories which is GR with all possible values of the cosmological constant within a single theory. 

The $\alpha'$-expansion on its own points toward a zero cosmological constant. However, once we include string loop corrections, the situation changes. We will revisit the Fischler-Susskind approach~\cite{Fischler1986,Fischler1986b,Fischler1987}towards including the lowest order string loop correction in the picture. In this way, an arbitrary cosmological constant is generated through the string-loop corrections in the EFT. In this way, the EFT that we need to write down within the GR EFT to include the string-loop corrections contains an arbitrary cosmological constant, which is exactly what happens with the UG EFT, although in the former case it is a coupling constant whereas in the latter it is an integration constant. In this way, we conclude that both the UG and the GR EFTs can account for the low-energy description of massless string states with the only difference arising in the nature of the cosmological constant. 

It is worth remarking that this analysis gives further evidence for UG as a sensible classical theory of gravitation according to the criteria invoked by Weinberg in~\cite{Weinberg1988}. According to Weinberg, one of the key aspects that needs to be addressed to regard UG as a reasonable classic theory of gravitation is to understand whether it can be obtained as a low energy limit of a quantum theory of gravitation. By embedding UG within the framework of string theory, here we answer here in the affirmative.

The remain of this article is structured as follows. In Section~\ref{Sec:UG_intro} we introduce the framework of UG, making special emphasis on the existence of a priviliged background volume form and the existence of an additional global degree of freedom with respect to GR. Then, we introduce a modification of GR in which a new global degree of freedom, precisely the cosmological constant is assigned to a $(D+1)$-form field, to make clear the difference between UG and the standard formulation of GR. In Sec.~\ref{Sec:Trivial_Background} we review the basics of the quantization of strings in flat spacetimes and explain why UG and GR are both valid as the low energy description of string theory from the point of view of scattering amplitudes involving massless particles. In Sec.~\ref{Sec:Background_String_Theory} we move on to analyze strings in general backgrounds. In Subsec.~\ref{Subsec:Betas} we rederive the consistency conditions (Weyl anomaly cancellation) from the perturbative $\alpha'$ expansion of the sigma model. Some of the details of the computation that are well explained in the literature and not relevant for our purposes are skipped and we refer the reader to the literature at those points. In Subsec.~\ref{Subsec:String_Loops} we introduce the Susskind-Fischler approach for cancelling some of the divergences arising from string loops, with the divergences of the sigma model on the trivial genus worldsheet. The main novelty that this mechanism introduces is a cosmological-constant-like term in the $\beta$ functions. We close this section by analyzing in Subsec.~\ref{Subsec:EFTs} how these consistency conditions can be derived from an effective action once they are interpreted as equations of motion for the background fields. We emphasize the consistency of this approach when computing scattering amplitudes involving the massless excitations. we close this section. In Sec.~\ref{Sec:Conclusions} we summarize the results and draw the conclusions that can be taken from our analysis. We also point interesting future lines of work that seem promising in virtue of our analysis presented here. 

\emph{Notation and conventions:} Our convention for the signature of the metric is $(-,+,...,+)$ for the $(D+1)$-dimensional target space metric and $(-,+)$ for the worldsheet metric. Tensor objects will be represented by bold symbols, whereas their components in a given basis will be written with the same (not bold) symbol and indices, e.g., the Minkowski metric $\boldsymbol{\eta}$ will be represented in components as $\eta_{\mu \nu}$. We will use Greek letters for spacetime indices $(\mu, \nu, ...)$ whereas we will reserve lower case latin indices $(a,b,...)$ for the worldsheet indices. Curvature quantities like the Riemann tensor are defined following Misner-Thorne-Wheeler's conventions~\cite{Misner1973} and we will specify explicitly the metric it depends on, e.g. $R^{\alpha}_{\ \beta  \gamma \delta} (\boldsymbol{g})$. We also represent the $(D+1)$-dimensional Newton's constant as $\kappa^2 = 16 \pi G$.

\section{Unimodular Gravity and General Relativity: Matching global degrees of freedom}
\label{Sec:UG_intro}

It is well accepted that metric theories of gravity, those in which the fundamental object describing the gravitational field at a given point is a metric, are suitable for describing gravitational experiments to great accuracy~\cite{Will2018}. The metric at a given point of the spacetime is completely specified by the lightcone at that point up to a conformal factor. Although the conformal structure of the spacetime is allowed to fluctuate both in UG and GR, the difference arises in the conformal factor. Whereas in UG the conformal factor is fixed to be a fiducial (non-dynamical) volume form that we represent as $\boldsymbol{\omega} = \frac{1}{(D+1)!} \omega(x) dx^{0} \wedge ... \wedge dx^{D}$ and hence it does not have any dynamics, in GR it is also dynamical like the lightcone itself. 

Naively, one could conclude that this reduction in the number of independent components of the metric may lead to a reduction of the independent degrees of freedom of the theory. However, it reduces the gauge symmetries of the theory to only transverse diffeomorphisms (those preserving the background volume form) and hence it is not surprising that the theory displays the same number of local degrees of freedom as GR does. Actually, it displays an additional global degree of freedom associated with the cosmological constant. In this section we will introduce the basic formulation of UG, emphasizing the presence of this new additional global degree of freedom. Furthermore, we will present a formulation of GR closer in spirit to UG, since the cosmological constant appears as a combination of an arbitrary integration constant and the renormalized cosmological constant entering the action and we still have the invariance under the full set of diffeomorphisms. 

Let us begin with the standard formulation of UG. UG is a theory in which the group of gauge transformations is WTDiff (Weyl rescalings of the metric and Transverse Diffeomorphisms) instead of the whole group of Diffs (Diffeomorphisms), see~\cite{Carballo-Rubio2022} for further details. In order to define such a theory, we need to use the non-dynamical volume form that we have already introduced $\boldsymbol{\omega}$. It is useful to introduce the Weyl-invariant auxiliary metric
\begin{align}
    \tilde{g}_{\mu \nu} = g_{\mu \nu} \left( \frac{\omega^2}{\abs{g}} \right)^{\frac{1}{D+1}}.
\end{align}
In this way, every curvature scalar built from the auxiliary metric $\tilde{g}_{\mu \nu}$ inherits the invariance under Weyl rescalings and is also invariant under transverse-diffeomorphism transformations by construction. The simplest action principle that one can think for an UG-like theory is the UG version of the Einstein-Hilbert action:
\begin{align}
    S_{\text{UG}} = \frac{1}{2 \kappa^2} \int d^{D+1} x \omega  R \left( \boldsymbol{g} \right).
\end{align}
We can also add a coupling to some matter fields which need to couple to the auxiliary metric, i.e., the matter action will be of the form $S_m \left( \boldsymbol{\tilde{g}}, \Phi \right)$, so that it remains Weyl-invariant (note that the matter fields are not affected by Weyl transformations). The equations of motion of this theory are the traceless Einstein equations: 
\begin{equation}
    R_{\mu \nu} (\boldsymbol{\tilde{g}}) - \frac{1}{D+1} R (\boldsymbol{\tilde{g}}) \tilde{g}_{\mu \nu } = \kappa^2 \left( T_{\mu \nu} (\boldsymbol{\tilde{g}}) - \frac{1}{D+1} T (\boldsymbol{\tilde{g}}) \tilde{g}_{\mu \nu} \right).
    \label{traceless_einstein}
\end{equation}
Upon using the Bianchi identities, they become Einstein equations with the cosmological constant entering as an integration constant~\cite{Carballo-Rubio2022}
\begin{equation}
    R_{\mu \nu} (\boldsymbol{\tilde{g}}) - \frac{1}{2} R (\boldsymbol{\tilde{g}}) \tilde{g}_{\mu \nu } + \Lambda \tilde{g}_{\mu \nu}= \kappa^2 T_{\mu \nu} (\boldsymbol{\tilde{g}}), 
\end{equation}
provided that $ \tilde{\nabla}_{\mu} T^{\mu \nu} \left( \boldsymbol{\Tilde{g}} \right) = 0$. 

It is clear that the Weyl invariance is trivial in the sense that its gauge fixing is trivial, we simply need to fix the volume form given by the determinant of the metric $\sqrt{\abs{g}}$ to be the background volume form. Actually, this can be done also at the level of the action. The result is still a local action for the metric which does not contain any mention to the Weyl symmetry. In that sense, the resulting action is the most minimalistic action that one can conceive for a metric field. If one tried to make a gauge fixing of the remaining degrees of freedom, one would end up with a non-local action for the actual physical degrees of freedom encoded in the field $g_{\mu \nu}$.

In this way, it seems clear that both theories display the same number of local degrees of freedom of GR, except for the cosmological constant that we will analyze now. To put it in other words, leaving aside the cosmological constant, from the point of view of initial conditions, the same amount of initial data are needed to specify a solution to the equations. The cosmological constant in this case appears with a difference, it is an additional global degree of freedom. The simplest way to see this is from the point of view of such constant being an integration constant. This means that it is a constant that parametrizes the space of solutions, which is separate from the initial data required in GR. In that sense, it is a constant to be fixed by initial conditions which makes the space of solutions of UG bigger than the GR space of solutions, precisely by this cosmological constant as an integration constant. This analysis can be made much more precise by making a Hamiltonian analysis of the theory, as it has been done in~\cite{Henneaux1989}, reaching the same conclusions.

We have concluded that UG is equivalent to GR, up to a global degree of freedom which is precisely playing the role of the cosmological constant. To make it more explicit, we will introduce now an additional field in GR that accounts for this global degree of freedom, to sharpen the difference. We need to introduce a $(D+1)$-form field which is the differential of a $D$-form~\cite{Henneaux1984,Mottola2022}. Explicitly, we want to introduce a $D+1$ form $\boldsymbol{F}$ which is the differential of a $D$-form $\boldsymbol{A}$. In components, this reads:
\begin{align}
    F_{\mu_0 ... \mu_D} = \nabla_{[\mu_0} A_{\mu_{\mu_1 ... \mu_D}]}. 
\end{align}
We can write down the action principle which is the Einstein-Hilbert action with an arbitrary cosmological constant and a Maxwell-like term for $\boldsymbol{F}$, namely:
\begin{align}
    S = \frac{1}{2 \kappa^2} \int d^{D+1} x \sqrt{-g} \left[- 2 \Lambda + R (\boldsymbol{g}) - \frac{K}{(D+1)!} F_{\mu_0 ... \mu_D} F^{\mu_0 ... \mu_D} \right],
\end{align}
where $K$ is simply a coupling constant which can be both positive or negative. The equations of motion for the $\boldsymbol{F}$-field are 
\begin{align}
    \nabla_{\mu_0} F^{\mu_0 ... \mu_D} = 0. 
\end{align}
In a $(D+1)$-dimensional manifold, a completely antisymmetric volume form like $\boldsymbol{F}$ needs to be proportional to the $\boldsymbol{\epsilon}$ pseudotensor. Hence, the equations of motion simply fixed the proportionality function to be a constant, i.e. 
\begin{align}
    F_{\mu_{0}...\mu_{D}} = c \sqrt{-g} \epsilon_{\mu_{0}...\mu_{D}}.
\end{align}
From the point of view of the initial value problem, this constant $c$ is precisely a global degree of freedom that needs to be fixed in terms of initial conditions. From that point of view, it is akin to the cosmological constant in UG, since it is completely fixed in terms of the initial conditions. We can sharpen the analogy by examining how does this constant $c$ enter the equations of motion for the metric. The energy-momentum tensor once we evaluate the $\textbf{F}$ form on shell, behaves exactly as a cosmological constant~\cite{Henneaux1984,Mottola2022}. Assuming the existence of additional matter fields, the equations of motion for the gravitational field take the following form:
\begin{align}
    R_{\mu \nu} (\boldsymbol{g}) - \frac{1}{2} R (\boldsymbol{g}) g_{\mu \nu } + \Lambda_{\text{eff}} g_{\mu \nu}= \kappa^2 T_{\mu \nu} (\boldsymbol{g}),
\end{align}
where the constant $\Lambda_{\text{eff}}$ is expressed in terms of the action as
\begin{align}
    \Lambda_{\text{eff}} = \Lambda + N_D K c^2,
\end{align}
with $N_D$ an irrelevant numerical factor depending on the spacetime dimension. In this way, the cosmological constant entering the equations of motion for the metric are a combination of an initial condition $c$ and the cosmological constant $\Lambda$ entering the action. 

From a purely classical point of view, we have presented a theory akin to GR, exhibiting the whole set of diffeomorphisms as gauge symmetries and containing an additional global degree of freedom encoded in a $(D+1)$-form. The equations of motion for this volume form enforce that it is proportional to the Levi-Civita pseudotensor, with the proportionality constant been called here $c$. The constant of proportionality enters the equations of motion for the metric as an effective cosmological constant. In this way, it plays a similar role to the one played by the global degree of freedom of UG. Independently of the value that we assign to the cosmological constant entering the action $\Lambda$, the resulting effective cosmological constant entering Einstein equations $\Lambda_{\text{eff}}$ is given by a combination of $\Lambda$ and $c$. In terms of the initial conditions, it is possible to adjust $c$ in order to make $\Lambda_{\text{eff}}$ take any desired value. This formulation of GR with the additional $(D+1)$ form field is equivalent to UG, in the sense that it displays the same amount of degrees of freedom, both local and global, and the global degree of freedom plays the role of a cosmological constant. 

At the quantum level, both formulations seem to be different from the point of view of radiative corrections. The reason behind this mismatch is that, whereas in UG the cosmological constant does not receive any radiative corrections and this makes it technically natural~\cite{Carballo-Rubio2015,Carballo-Rubio2022}\footnote{We note that technical naturalness is a definition that only applies to coupling constants appearing in the action. In that sense it is not completely legitimate to say that in UG the cosmological constant is technically natural since it is not a coupling constant. However, making an abuse of language we find it convenient to say that it is technically natural.}, in this formulation of GR, the cosmological constant in the action $\Lambda$ does receive radiative corrections, and hence it is not technically natural. However, the cosmological constant relevant for the dynamics is the effective one $\Lambda_{\text{eff}}$ that combines the renormalized $\Lambda$ with the initial value constant $c$. It is possible to obtain any value for the cosmological constant $\Lambda_{\text{eff}}$ independently of the potentially huge radiative corrections that $\Lambda$ may receive. The equivalence once quantum corrections are included into the picture is unclear. Whether this formulation is then completely equivalent to UG at the semiclassical level is something that deserves a separate and detailed study on its own. 

Our point here was mainly to introduce a formulation within the GR setup that is close to the UG version, so that both theories can be compared easily. We have made explicit the difference existing in the global degrees of freedom of UG and GR (UG contains the whole space of GR with arbitrary values of the cosmological constant coupling). This only difference in the two theories, will be also the only difference appearing from the point of view of regarding UG as the low energy EFT for massless string states.

\section{String perturbation theory in trivial backgrounds}
\label{Sec:Trivial_Background}

This section contains a review of the quantization of strings in a flat background as well as the computation of string scattering amplitudes for gravitons from string theory. This is well-known material that can be found in any textbook~\cite{Polchinski1998a,Green1987a}. Also we think that a reader unfamiliarized with string theory might find here a quick introduction to the arguments presented in the literature leading to the conclusion that GR is the EFT describing the excitation in massless degrees of freedom. We find convenient to make such introduction here to expand the discussion presented in~\cite{Carballo-Rubio2022} about how the scattering amplitudes can be equivalently obtained from a GR and a UG-like EFT. 

The starting point of our discussion of perturbative string theory is the action describing relativistic strings propagating in flat spacetime. For relativistic free particles it is natural to consider the action to be the proper time of the particle trajectory i.e., the embedding of the worldline in the target space. In the same way, for strings it is natural to consider the area swept out by the worldsheet to replace the proper time of the particle trajectory. For that purpose, let us introduce a coordinate system in the worldsheet, a pair $\sigma^a$ ($a = 0,1$) which correspond to the time coordinate $\sigma^0 \in (-\infty,\infty)$ and a spatial coordinate $\sigma^1$. Furthermore, we will restrict our attention to closed strings (those giving rise to graviton excitations) in which the points at $\sigma^1$ and $\sigma^1 + 2 \pi$ are identified. If we endow the $(D+1)$ dimensional flat spacetime with coordinates $X^{\mu}$, we look for an action such that the area density swept by the string is expressed in terms of derivatives of the embedding $X^{\mu} (\tau, \sigma)$. We notice that the induced metric on the worldsheet is given by 
\begin{align}
    h_{a b} = \eta_{\mu \nu} \partial_{a} X^{\mu} \partial_{b} X^{\nu}.
\end{align}
If we take the action to be the area swept out by the string, we write down the Nambu-Goto action as 
\begin{align}
    S_{NG} [X] = - \frac{1}{2 \pi \alpha'} \int d^2 \sigma \sqrt{-h}.
\end{align}
The constant $\alpha'$ represents the string tension, i.e., the energy density per unit length. Although this action is perfectly reasonable classically, from the point of view of quantization is problematic. This is because it is not quadratic in its variables: we have a square root appearing explicitly in the action. To circumvent this problem, one can work with the Polyakov action, which is given by 
\begin{align}
    S_P[X,\gamma] = -\frac{1}{4 \pi \alpha'} \int d^2 \sigma \sqrt{-\gamma} \gamma^{a b} \partial_{a } X^{\mu} \partial_{b} X^{\nu} \eta_{\mu \nu}. 
\end{align}
In this action, there is an additional configuration variable $\gamma_{ab}$ which is a metric in the worldsheet. Now, this action is clearly quadratic in the $X^{\mu}$ variables over which we will path-integrate to quantize the theory. To see the equivalence among these two actions, we can compute the equations of motion for the $\gamma_{ab}$ variable. Actually, following the standard conventions, we can define a two-dimensional energy-momentum tensor as the variation of the Polyakov action with respect to the worldsheet metric, i.e. $\gamma_{ab}$: 
\begin{align}
    T_{a b} = -  \frac{1}{\sqrt{-\gamma}} \frac{\delta S_P}{\delta \gamma^{ab}} = \frac{1}{4 \pi \alpha'} \left[ \partial_{a} X^{\mu} \partial_b X^{\nu} - \frac{1}{2} \gamma_{a b} \gamma^{a d}  \partial_{c} X^{\mu }\partial_{d} X^{\nu} \right]\eta_{\mu \nu}.
\end{align}
The Polyakov action does not contain any derivatives of the metric $\gamma_{a b}$, and hence the equations of motion for the metric can be regarded as a constraint $T_{a b} = 0$ (as a consequence, strictly speaking it is not a dynamical variable). Actually, this constraint can be used to solve $\gamma_{a b}$ in terms of the $X^{\mu}$ variables. When we plug the result back into the Polyakov action, we find the Nambu-Goto action we began with. 

It is worth pausing at this point and discussing the continuous symmetries of the theory:
\begin{itemize}
    \item Poincar\'e invariance. This is a global symmetry on the worldsheet 
    \begin{align}
        & X^{\mu} \rightarrow \Lambda^{\mu}_{\ \nu} X^{\nu} + c^{\mu}.
    \end{align}
    
    \item Reparametrization invariance or diffeomorphism invariance in the worldsheet $  \sigma^a \rightarrow \Tilde{\sigma}^a (\sigma) $. Whereas the $X^{\mu}$ fields transform as worldsheet scalars, $\gamma_{ab}$ transforms as a two-index covariant tensor: 
    \begin{align}
        & X^{\mu} (\sigma) \rightarrow X^{\mu} (\tilde{\sigma}) = X^{\mu} (\sigma) , \\
        & \gamma_{a b} (\sigma) \rightarrow \tilde{\gamma}_{ab} (\tilde{\sigma}) = \frac{\partial \sigma^c}{ \partial \tilde{\sigma}^a} \frac{\partial \sigma^d}{ \partial \tilde{\sigma}^b} \gamma_{c d} (\sigma).
    \end{align}
    \item Weyl invariance of the worldsheet metric $\gamma_{ab}$. This transformation leaves invariant the $X^{\mu}$ coordinates and the metric gets a local rescaling 
    \begin{align}
        & X^{\mu} (\sigma) \rightarrow X^{\mu} (\sigma) , \\
        & \gamma_{a b} (\sigma) \rightarrow e^{2 \phi (\sigma) } \gamma_{a b} (\sigma).
    \end{align}
\end{itemize}
We can distinguish now between oriented and unoriented strings. The former have a well defined transformation law under the parity transformation $\sigma^1 \rightarrow 2 \pi - \sigma^1$. We will focus on the unoriented strings for the sake of simplicity. 

Not all the symmetries that we have introduced are 
directly preserved through the process of quantization. Actually, the Weyl symmetry is anomalous, as it is well known. However, in this case the Weyl symmetry is a gauge symmetry that we must insist on preserving at the quantum level to remove unphysical states. We will further discuss this point later when we deal with strings in general backgrounds. For the time being, let us focus on the quantization of the theory through a path-integral procedure. 

Let us illustrate the quantization of the theory through a path-integral procedure as well as the spectrum that the theory displays. Let us define the generating functional following the usual Faddeev-Popov procedure. First of all, we would write down the action in Euclidean space, in order to make the quantization procedure sensible. We write down the generating functional as
\begin{equation}
    Z = \frac{1}{V(\text{gauge})} \int \mathcal{D} \gamma \mathcal{D} X e^{-S_P [X,\gamma]},
\end{equation}
where $V(\text{gauge})$ represents the volume of the gauge group. We recall that we have the Weyl rescalings of the metric and diffeomorphisms as gauge symmetries of our theory. Hence, we need to avoid counting more than once physical configurations and that is the reason for taking the quotient by the volume of the gauge group. As usual, we will introduce a Faddeev-Popov determinant $\Delta_{FP} [\gamma]$ to take this volume into account.

The integral over the gauge orbits cancels with the volume of the gauge group and we reach the expression for the generating functional which is
\begin{align}
    Z[\gamma]  =  \int  \mathcal{D} X \Delta_{FP} [\gamma ]  e^{-S_P [X,\gamma]}. 
\end{align}
Choosing a convenient normalization for the action, we can rewrite the Faddeev-Popov determinant as 
\begin{align}
    \Delta_{FP} [\gamma] = \int \mathcal{D} b \mathcal{D} c e^{- S_{\text{g}} [b,c]},
\end{align}
where $b$ and $c$ are ghosts Grassman-values that anticommute and 
\begin{align}
    S_{\text{g}} = \frac{1}{2 \pi} \int d^2 \sigma \sqrt{\gamma} b_{a b} \nabla^a c^b.
\end{align}
At this point, we have reduced the evaluation of the path integral for the bosonic string theory to the evaluation of the path integral: 
\begin{align}
    Z = \int \mathcal{D}b \mathcal{D}c \mathcal{D}X e^{- S_{P}[\gamma,X] - S_{\text{g}} [\gamma, b, c]}, 
\end{align}
which is the conformal field theory (CFT) of $D+1$ scalar fields (the $X^{\mu}$) and the $bc$-ghost system~\cite{Polchinski1998a,Nakayama2013}. If the theory is going to preserve the Weyl invariance, we need the theory to have a total zero central charge. This is precisely the consistency condition that we mentioned would appear. Weyl invariance means that the trace of the two-dimensional energy momentum tensor needs to vanish. In two-dimensions, the trace of the energy-momentum tensor is determined by the central charge and the trace anomaly 
\begin{align}
    \expval{T^a_{\ a}} = - \frac{c}{12} R \left[ \gamma \right].
\end{align}
The system of the $X^{\mu}$-scalars and the $bc$-ghost system is linear, and hence the total central charge is the sum of the central charges of the two systems independently:
\begin{align}
    c = c_{g} + c_{X}.
\end{align}
The $bc$-ghost system~\cite{Polchinski1998a} has a central charge $c_{g} = - 26$ while each scalar field gives a contribution of $1$ to the central charge $c_{X} = D + 1 $. Ensuring Weyl-invariance means that we need the spacetime dimension to be $26$. This is the well-known way in which the critical dimension of bosonic string theory emerges.

Now that we have ensured how to preserve the gauge invariance at the quantum level in order to make the theory consistent, it is time to talk about the spectrum of the strings. Our point is simply to illustrate that the spectrum of the closed unoriented bosonic string contains a tachyon, a dilaton, and a graviton. Hence, for this purpose, we can skip the detailed BRST analysis and focus only on the states generated by the $X$-fields which are the ``physical fields''. 

In order to characterize the spectrum, the simplest way to do it is to use the so called state-operator map for CFTs~\cite{Qualls2015,DiFrancesco1997}, in which states are replaced by operator insertions that generate them by acting in a neighbourhood of the vacuum. For this purpose, it is first easier to use complex coordinates $\sigma \rightarrow (z,\bar{z})$ on the worldsheet. Furthermore, we now need the operators to be gauge invariant. The diffeomorphism invariance can be ensured by integrating local operators $\mathcal{O}(z, \bar{z})$ over the worldsheet, i.e. constructing operators of the form
\begin{align}
    V = \int d^2z \mathcal{O} (z, \bar{z}), 
\end{align}
with $V$ standing for vertex operators. Weyl invariance is ensured by choosing the operators $\mathcal{O}$ to transform adequately under Weyl rescalings, i.e., having a suitable weight. The measure of integration, $d^2z$ has a conformal weight $(-1,-1)$ under such rescalings. Hence, $\mathcal{O}$ needs to be a primary operator of the CFT with weight $(+1,+1)$ to compensate it. 

The kind of operators that give rise to the lowest energy states of the string are $e^{i p \cdot X}$ and $P_{\mu \nu} \partial X^{\mu} \partial X^{\nu} e^{i p \cdot X}$, with $p$ a given momentum that we endow the string with and $P_{\mu \nu}$ the polarization tensor~\cite{Green1987a,Polchinski1998a}. The operator $e^{i p \cdot X}$ gives rise to the tachyon, since we need to impose that $ - p^2 = - 4/\alpha' < 0$ for the operator to be Weyl invariant. The operator $P_{\mu \nu} \partial X^{\mu} \partial X^{\nu} e^{i p \cdot X}$ corresponds to the dilaton (pure trace part of $P_{\mu \nu}$) and the symmetric part of $P_{\mu \nu}$ gives rise to the graviton, since $p^2 = 0$ (massless condition) and $p^{\mu} P_{\mu \nu} = 0$ (transverse condition) needs to be imposed to ensure the Weyl invariance. The antisymmetric part does not appear for unoriented strings since it corresponds to the Kalb-Ramond excitation~\cite{Polchinski1998a}.

Up to this point, we have analyzed the spectrum of the closed unoriented bosonic string theory and found that the massless states correspond to the dilaton and the graviton. The Polyakov action \emph{per se} does not give rise to interactions. We will now make a small digression on how interactions among the massless states arise in string theory. 
There is a term that we can add to the Polyakov action which is an Einstein-Hilbert term that is purely topological in two-dimensions
\begin{align}
    S_{\text{int}} = \frac{\lambda}{4 \pi} \int d^2 \sigma \sqrt{\gamma} R(\gamma) = 2 \lambda (1-g), 
\label{Eq:Int_Action}
\end{align}
being $g$ the genus of the worldsheet and $\lambda$ a coupling constant which we assume to be small in order to do perturbation theory. Hence, if we add this term to the string action, we will get
\begin{align}
    Z = \sum_{\text{topologies}} \int \mathcal{D} X \mathcal{D} \gamma e^{-S_P -  S_{\text{int}}} = \sum_{g=0}^{\infty} e^{-2\lambda (1-g)} \int \mathcal{D} X \mathcal{D} \gamma e^{-S_P}.
\end{align}
If we call $e^{\lambda} = g_s$, as it is common, this gives a good expansion as long as $g_s \ll 1$. The whole series is known to be a divergent series as the standard perturbative series in QFT~\cite{Gross1988}. In addition to this problem, there is a harder problem which is the finiteness of \emph{each} of the terms in the series, i.e., the path integral over the different geometries. For a fixed topology, the path integral with the Polyakov action requires to compute a sum over the moduli of conformally inequivalent surfaces. In general, for higher loop orders (i.e. non-trivial topologies) this requires to perform an integral over a moduli space that is not obviously convergent, although some results in the literature point toward its finiteness~\cite{Mandelstam1991}.  

Now it comes to the point of computing some observables. The observable to compute in string theory is the string $S$-matrix. This means, we plug some ``in" state of the free string spectrum and compute the probability amplitude of generating another ``out" state of free string spectrum. These states are generated by introducing their corresponding vertex operators.

For our purposes of analyzing how GR or UG might emerge from string theory, we are interested in computing the scattering amplitude involving $m$ gravitons with momenta $\boldsymbol{p}_i$ and polarization tensors $\boldsymbol{e}_i$ which we represent as $\mathcal{A}^{(m)} (p_1,e_1; p_2,e_2; ... p_m, e_m)$. This is computed as a suitable path integral for the Polyakov action $S_P$ that schematically reads~\cite{Green1987a,Polchinski1998a}
\begin{equation}
    \mathcal{A}^{(m)} (1^{h_1},2^{h_2},...,m^{h_m}) = \frac{1}{g_s^2} \frac{1}{V_\textrm{gauge}} \int \mathcal{D} X \mathcal{D} \boldsymbol{g} \,e^{- S_\text{P} [X,\boldsymbol{g}]} \prod_{i=1}^{m} V_{i} (p_i, h_i),
\end{equation}
where $V_i$ represents the vertex operator associated with a graviton insertion with a given spin and momentum. To begin with, we particularize the amplitude for three gravitons and we find 
\begin{equation}
    \mathcal{A} (p_1,e_1; p_2, e_2; p_3, e_3) = i \frac{g_s (\alpha^{\prime})^6}{2}(2 \pi)^{26} \delta^{26} \left( p_1 + p_2 + p_3 \right) e_{1 \mu \nu} e_{2 \alpha \beta} e_{3 \gamma \delta} T^{\mu \alpha \gamma} T^{\nu \beta \delta},
\end{equation} 
where
\begin{equation}
    T^{\mu \alpha \gamma} = p_{23}^{\mu} \eta^{\alpha \gamma} + p_{31}^{\alpha} \eta^{ \gamma \mu } + p_{12}^{\gamma} \eta^{\mu \alpha} + \frac{\alpha'}{8} p^{\mu}_{23} p^{\alpha}_{31} p^{\gamma}_{12},
\end{equation}
$p_{ij}^{\mu} = p^{\mu}_i - p^{\mu}_{j}$. The terms of order $ \order{\alpha'}$ in $T^{\mu \alpha \gamma}$ contribute as $\order{p^4}$ to the amplitude. If we focus just on the lowest order terms $\order{p^2}$, this amplitude is equivalent to the ones computed at tree level from the Einstein-Hilbert action upon the identification $\kappa = g_s ( \alpha')^6$. The same agreement is found with amplitudes involving an arbitrary number of gravitons: if we neglect the higher-order contribution from the string amplitude, they agree with those computed from the Einstein-Hilbert action~\cite{Polchinski1998a,Green1987a}, with the same identification of $\kappa$ and the string constants. 

As it has been already discussed in the literature~\cite{Alvarez2016,Carballo-Rubio2019}, the tree-level scattering amplitudes of gravitons computed in GR and UG are identical. Hence, from the point of view of scattering amplitudes, string theory does not point toward GR in a univocal way: both UG and GR are equivalent from a low-energy effective field theory point of view. This result was already advanced in~\cite{Carballo-Rubio2022} and we have reproduced here the analysis in more detail for the sake of completeness. We will come back to this analysis later, when we introduce the low energy EFTs for the massless degrees of freedom of the string: both the UG and the GR-like actions.  

\section{Strings in general backgrounds}
\label{Sec:Background_String_Theory}

Up to now, we have only considered strings propagating in flat spacetime. However, the spectrum of the strings contains some excitations which typically interact among themselves and could lead to the generation of a non-trivial background. In particular, it contains a graviton and, necessarily, gravitons need to interact gravitationally. At low energies, all the excitations that matter are the massless ones. In the same way a laser is a coherent state of photons, we expect that a coherent state of gravitons might look like a curved background and a string propagating on top of it needs to be described appropiately. The same comment applies to the dilaton field. As such, we can write down the most general renormalizable action including those fields, which is the following non-linear $\sigma$-model
\begin{align}
    S[X, \gamma] = S_P[X,\gamma] + S_D[X,\gamma]= -\frac{1}{4 \pi \alpha'} \int d^2 \sigma \sqrt{-\gamma} \left[ \gamma^{a b} \mathcal{G}_{\mu \nu} (X) \partial_{a } X^{\mu} \partial_{b} X^{\nu} + \alpha' R \left( \gamma \right) \Phi (X) \right],
\label{Eq:Action_String}
\end{align}
where $\mathcal{G}_{\mu \nu} (X)$ represents a metric (graviton excitations), $\Phi(X)$ represents the dilaton background field, and $R[\gamma]$ represents the Ricci-scalar of the two-dimensional metric. This term breaks explicitly the Weyl invariance in the worldsheet. This term is of a higher dimension than the Weyl-invariant terms, and it does not require to be normalized with a dimensionful constant. In virtue of the expansion in $\alpha'$ that we will perform, we will cancel the tree-level contribution to the anomaly of this last term with the one-loop contribution of the classically Weyl-invariant terms. The result of this procedure is a reasonable effective field theory for the massless degrees of freedom of the string.

There are two missing terms that still give rise to a renormalizable theory. The first of these terms is the coupling to the Kalb-Ramond field. However, if we focus on unoriented strings, we can skip it since the divergences of the rest of the terms do not require this term to be renormalized. In case we deal with oriented strings, this term gives a contribution to the conformal anomaly~\cite{Polchinski1998a}.

The additional term that we can add to the action corresponds to a coupling to the background tachyon field $T(X)$
\begin{align}
    S_T = \frac{1}{4\pi} \int d^2 \sigma \sqrt{- \gamma} T \left( X \right).
\end{align}
In principle this term is needed to cancel some of the quadratic divergences arising from vacuum to vacuum diagrams. However, if we use a renormalization scheme such that those divergences are absent (for example, dimensional regularization), we can safely skip those terms. Hence, we will work with a renormalization scheme fullfilling this property. Furthermore, it is worth mentioning that supersymmetry in the worldsheet ensures that those quadratic divergences are absent in superstrings due to the characteristic cancellation among fermionic and bosonic degrees of freedom, with independence of the renormalization scheme. 

\subsection{Determination of the Weyl anomaly}
\label{Subsec:Betas}

Anomalies always appear when there are two symmetries that the theory displays at the classical level, but it is not possible to quantize such theory preserving both of them. This means, there is a trade-off between the two symmetries and it is only possible to preserve one of them in the process. For example, the chiral anomaly is a trade-off between the vector and axial currents for massless fermion fields. If we use a regularization procedure which automatically preserves one of those currents, then straightforwardly the other current will be anomalous. In the case of the chiral anomaly, it is standard to use a regularization scheme that preserves gauge invariance and hence yields to the conservation of the vector current, leading to an anomalous axial current. In the case of Weyl invariance for strings, we are using a regularization scheme that preserves diffeomorphism invariance, while the Weyl symmetry becomes potentially anomalous. We need to ensure that the non-linear sigma model is chosen in such a way that it gives rise to a Weyl-invariant theory. In a language closer to particle physics, this means that we need to choose our theory in such a way that we cancel the potential gauge anomalies, which in this case corresponds to choosing the background fields in such a way that the theory is not Weyl-anomalous. In the case of the Standard Model, since it corresponds to a chiral gauge theory, arbitrary matter fields would lead to an anomalous theory. However, the matter content is such that the potential anomaly is absent. This is precisely what we have done in the previous section to fix the target space dimension to be $26$; otherwise, the Weyl-symmetry becomes anomalous. In this case, we expect constraints also on the background fields entering the non-linear sigma models, i.e., constraints that the $\mathcal{G}_{\mu \nu} (X)$ and the $\Phi(X)$ fields need to obey. 

We want now to write down the most general form that the Weyl anomaly can display. Following D'Hoker~\cite{Deligne1999}, it is possible to show that the structure of the anomaly for unoriented strings in a curved background needs to be of the form
\begin{align}
    \expval{T_{a}^{\ a}} = \beta_{\mu \nu}^{G}(X) \partial_{a} X^{\mu} \partial_{b} X^{\nu} \gamma^{ab}+ \beta^{\Phi} \left( X \right) R \left( \gamma \right)  + \beta^V_{\mu}(X) g^{a b} D^{\ast}_{a} \partial_{b} X^{\mu},
\label{Eq:Trace_Anomaly}
\end{align}
where $D^{\ast}_a$ represents the covariant derivative on the product space of the cotangent space of the worldsheet and the tangent space of the target space, and it can be explicitly written down as  
\begin{align}
    D^{\ast}_a \partial_b X^{\mu} = \partial_a \partial_b  X^{\mu} - \Gamma^{c}_{\ ab} \partial_c X^{\mu}+ \Gamma^{\mu}_{\ \nu \rho} \partial_{a} X^{\nu} \partial_b X^{\rho}, 
\end{align}
where $\Gamma^{c}_{\ ab}$ are the Christoffel symbols of the metric $\gamma_{a b}$ and $\Gamma^{\mu}_{\ \nu \rho}$ represent the Christoffel symbols of the metric $G_{\mu \nu}$. The last term in the Weyl anomaly, $\beta^V$ can be removed through a transformation on the $X^{\mu}$ fields, since we are always able to perform a local transformation on the $X^{\mu}$ fields at the same time that we perform a Weyl-rescaling of the metric. This leaves only two independent $\beta$ functionals: $\beta^G$ and $\beta^{\Phi}$\footnote{For oriented strings there will be another $\beta$-functional associated with the Kalb-Ramond field.}.

Hence we need to determine the $\beta$ functionals obtained from the action~\eqref{Eq:Action_String}. We want to study perturbatively this action order by order in the $\alpha'$ expansion, which is done by assuming that the background fields $G_{\mu \nu} (X),\Phi(X)$ vary smoothly with respect to the scale $\alpha'$. It is conventional to do the computations in the background field formalism. In this formalism, we decompose the fields $X^{\mu}$ in a background part $X_0^{\mu}$ and its quantum fluctuations $Y^{\mu}$
\begin{equation}
    X^{\mu} \left( \sigma \right) = X_0^{\mu} (\sigma) + Y^{\mu} \left( \sigma \right),
\end{equation}
where the integration is now performed with respect to the quantum fluctuations instead of $X^{\mu}$. We define the effective action $\Gamma [ X_0,g ] $ following~\cite{Callan1989} as 
\begin{align}
    e^{- \Gamma [ X_0,g ] } = \int \mathcal{D} Y e^{- \left[ S (X_0,Y)- S(X_0) - \int d^2 \sigma Y^{\mu} (\sigma) \frac{\delta S}{\delta X_0^{\mu}} \right] },
\end{align}
which is the generating functional of the Feynman diagrams relevant for the computation of the $\beta$-functionals. 

At this point, it is better to pause and mention a crucial step in the computations. The coordinate difference does not transform in a covariant way under changes of coordinates. Hence, in order to obtain results that are manifestly covariant, it is better to do the computation in variables that are manifestly covariant at intermediate steps. This can be done as follows. Imagine that the coordinates $X^{\mu}_0$ correspond to a given point $p_0$ and the coordinates $X^{\mu} = X^{\mu}_0 + Y^{\mu}$ to a point $p$. If both points are close enough, there exists only one geodesic with respect to $\mathcal{G}_{\mu \nu}$ connecting both of them. Hence, we can replace the coordinate difference $Y^{\mu}$ which characterizes the point $p$ by the tangent vector $t^{\mu}$ of the geodesic at the point $p_0$, which transforms covariantly under changes of coordinates. Hence, it is better to use this vector as the integration variable in the path integral. 

In fact, we can use this tangent vector $t^{\mu}$ to perform a covariant Taylor expansion based on $X_0^{\mu}$ of an arbitrary tensor living in the target manifold. To put it explicitly, any tensor $T_{\mu_1 \hdots \mu_n} (X)$ can be expanded as
\begin{align}
    T_{\mu_1 \hdots \mu_n} (X_0 + t) = \sum_{k=0}^{\infty} T^{(k)}_{\mu_1 ... \mu_n \nu_1 ... \nu_k} (X_0) t^{\nu_1 } \hdots t^{\nu_k},
\end{align}
where each of the terms $T^{(k)}_{\mu_1 ... \mu_n \nu_1 ... \nu_k} $ is a combination of covariant derivatives of the tensor $T_{\mu_1 \hdots \mu_n} $ and contractions with curvature tensors evaluated at $X_0$. This expansion can be achieved with the help of the normal coordinate expansion although we emphasize that it remains valid in an arbitrary coordinate system since it is a tensor expression. We are interested in the expansion of the tensors $\mathcal{G}_{\mu \nu}, \Phi(X)$ (the latter is a trivial tensor, i.e. a scalar), and the object $\partial_a \left( X_0^{\mu} + Y^{\mu} \right)$. These expansions can be obtained after a straightforward computation, see~\cite{Callan1989} for details:
\begin{align}
    \mathcal{G}_{\mu \nu} (X) & = \mathcal{G}_{\mu \nu} (X_0) + \frac{1}{3} \mathcal{R}_{\mu \rho \sigma \nu} t^{\rho} t^{\sigma } + ..., \\
    \Phi(X) & = \Phi(X_0) + \nabla_\mu \Phi(X_0) t^{\mu} + \frac{1}{2} \nabla_\mu \nabla_\nu \Phi(X_0) t^{\mu} t^{\nu} + ..., \\
    \partial_a \left( X^{\mu}_0 + Y^\mu\right) & = \partial_a X^{\mu}_0 + \nabla_a t^{\mu} + \frac{1}{3} \mathcal{R}^{\mu}_{\ \nu \rho \sigma} \partial_a X^{\sigma}_0 t^{\nu} t^{\rho} + ..., 
\end{align}
where $\mathcal{R}^{\mu}_{\ \nu \rho \sigma}$ represents the Riemann tensor associated with $\mathcal{G}_{\mu \nu}$. 

We are not ready to perform the diagrammatic computation yet. There is a problem arising from the fact that the term that gives us the propagator for the quantum fields over which we integrate, $t^{\mu}$, contains an arbitrary metric in front of it, i.e. we need to invert a term that looks like $\mathcal{G}_{\mu \nu} (X_0) \nabla_a t^{\mu} \nabla_b t^{\nu}$. The way to deal with this problem and obtain a simple propagator is to introduce a vielbein $e^{A}_{\ \mu} (X_0)$ which fulfills the property 
\begin{align}
    e^{A}_{\ \mu} (X_0) e^{B}_{\ \nu} (X_0) \eta_{A B} = \mathcal{G}_{\mu \nu} (X_0),
\end{align}
with $\eta_{AB}$ a Lorentzian metric. In this way, we can rewrite all the vector expressions in the non-holonomic basis $e^{A}_{\ \mu}$ and get a trivial propagator for the $t^{A}= e^{A}_{\ \mu} t^{\mu}$ fields. This comes with a subtlety, because now the derivatives $\nabla_a$ involve the spin-connection of the spacetime $\omega_{\mu}^{\ AB}$; for example, 
\begin{align}
    \nabla_a t^A = \partial_a t^A + \omega_{\mu}^{\ AB} \partial_a X^{\mu}_0 t^{C} \eta_{BC}.
\end{align}
Obtaining a trivial propagator means breaking the $SO(D,1)$ invariance that the theory displays, but since we are working in a formalism that is explicitly gauge covariant, we automatically know that there will always be contributions in the diagrammatic expansion that make the theory explicitly gauge covariant in intermediate steps. Up to this point, collecting all the information, we have performed the following expansion for the Polyakov piece of the action:
\begin{align}
    S_{P} & = S_P[X_0] + \frac{1}{2\pi \alpha'} \int d^2 \sigma \sqrt{\gamma} \gamma^{ab} G_{\mu \nu} (X_0) \partial_a X_0^{\mu} \nabla_b t^{\nu} \\
    & + \frac{1}{4 \pi \alpha'} \int d^2 \sigma \sqrt{\gamma} \gamma^{ab} \left[ \eta_{A B} \nabla_a t^A \nabla_b t^B \right] \\
    & + \frac{1}{3 \pi \alpha'} \int d^2 \sigma \sqrt{\gamma} \gamma^{ab} R_{\mu A B C} \partial_a X_0^{\mu} t^A t^B \nabla_{b} t^C \\
    & + \frac{1}{12 \pi \alpha'} \int d^2 \sigma \sqrt{\gamma} \gamma^{ab} R_{A B C D} t^{B} t^{C} \nabla_{a} \nabla_{a} t^{A}  \nabla_{b} t^{D},
\end{align}
and for the dilaton part we have the trivial structure: 
\begin{align}
    S_D [X_0 + t] = & S_D[X_0] - \frac{1}{8 \pi} \int d^2 \sigma \sqrt{\gamma}  \nabla_A  \Phi(X_0) t^A \\
    & - \frac{1}{16 \pi} \int d^2 \sigma \sqrt{\gamma}  \nabla_A \nabla_B \Phi(X_0) t^A t^B + ... \ . 
\end{align}
We recall that we can safely impose the equations of motion for the classical fields and safely drop the linear terms. This is tantamount to a legitimate field redefinition.

Now we can determine the trace anomaly, see Eq.~\eqref{Eq:Trace_Anomaly} from the effective action introduced above. The computation requires to go to the next higher order in loops in the dilaton field, since the piece of the action for the dilaton field $\alpha'$ comes with an additional $\alpha'$ with respect to the other field. The computation is rather lengthy and hence we do not reproduce it here~\cite{Callan1989}. We simply write down the result as 
\begin{align}
    & \beta^G_{\mu \nu} =  R_{\mu \nu} \left( \boldsymbol{G} \right)- \nabla_{\mu} \nabla_{\nu} \Phi + \order{\alpha'}, \\ 
    & \beta^{\Phi} = \frac{D-26}{6} + \alpha' \left[ - R \left( \boldsymbol{G} \right) + 2 \nabla^2 \Phi +(\nabla \Phi)^2 \right] +  \order{\alpha^{\prime 2}}.
\end{align}
A comment is in order now. If we are dealing with a flat worldsheet, the vanishing of $\beta^G$ is enough to ensure the Weyl invariance at the quantum level, as long as we are working in $ D = 26$ dimensions, the critical dimension (see Eq.~\eqref{Eq:Trace_Anomaly}. Hence, in principle, we expect that the same applies to non-flat worldsheets, i.e. that the condition $\beta^{\Phi} = 0$ is not independent of $\beta^G = 0$. Actually, we have a non-trivial constraint coming from the Bianchi identity
\begin{align}
    & \nabla^{\mu} \left( R_{\mu \nu} \left( \boldsymbol{G} \right) - \frac{1}{2} R \left( \boldsymbol{G} \right) G^{\mu \nu} \right) = 0.
\end{align}
This ensures that we have to the computed order the following identity whenever $\beta^G_{\mu \nu} = 0$
\begin{align}
    \nabla^{\mu} \beta^G_{\mu \nu} =  \nabla_{\nu} \beta^{\Phi} = 0, 
\end{align}
as can be seen by direct calculation. This implies that $\beta^{\Phi}$ is a constant as long as $\beta^G =  0$. By continuity, this automatically implies at this level that $\beta^{\Phi} = 0$ for $D=26$~\cite{Callan1989}. From now on we will restrict ourselves to work in $D=26$ and make a comment on strings on non-critical dimension later.

\subsection{Including string-loop corrections}
\label{Subsec:String_Loops}

At this point, we have only focused on the zeroth-order in the $g_s$-expansion. Although it is clear that string loops should modify the results, it is not completely clear how those corrections must be included. One of the most accepted proposals is the Fischler-Susskind approach~\cite{Fischler1986,Fischler1986b,Fischler1987}. The idea behind such mechanism is that string loop divergences can be absorbed through a renormalization of the background fields in the non-linear sigma models. Let us illustrate this explicitly for unoriented closed bosonic strings. For the purpose of this section, it is simpler to work with a sharp cut-off as regularization scheme. 

The divergences in string loops appear when we have to sum over conformally inequivalent surfaces of a fixed topology (i.e. genus). For a fixed but arbitrary topology (i.e. we focus here on non-trivial topologies), this sum is an integral over a finite-dimensional parameter space, the so-called Teichm\"uller space~\cite{Polchinski1998a,Green1987a}. These integrals are divergent, but these divergences arise from handles that shrink to zero size. These divergences are equivalent to the divergences coming from inserting a local operator on the trivial-genus worldsheet. In a flat spacetime, the divergence appearing for the torus topology can be eliminated through the insertion of an operator $\frac{\log \Lambda}{2 \pi} \gamma^{ab} \eta_{\mu \nu}  \partial_a X^{\mu}\partial_b X^{\nu}$, with a suitable coefficient. Here $\Lambda$ is a suitable cut-off in the Teichm\"uller space. 

If we move to a curved geometry $G_{\mu \nu}$ with a non-trivial zero mode of the dilaton field~$\Phi$, we need to substitute the metric $G_{\mu \nu}$ and include a relative factor $e^{- \Phi}$ to account for the dependence of the path integral on the topology of the surface. We recall that the asymptotic value of the dilaton field $\lambda = \expval{\Phi}$ is identified with the string coupling constant $g_s = e^{\lambda}$ through an exponential relation, as it can be seen by comparison of the actions in Eq.~\eqref{Eq:Action_String} and Eq.~\eqref{Eq:Int_Action}~\cite{Polchinski1998a,Green1987a}. Explicitly for the first non-trivial order (torus topology) we have the following divergences: 
\begin{align}
    \delta S^{\text{loop}} = \frac{\log \Lambda}{2 \pi} \int d^2 \sigma \sqrt{-\gamma} \gamma^{ab} e^{-\Phi} G_{\mu \nu} (X) \partial_a X^{\mu}\partial_b X^{\nu}.
\end{align}
The $e^{- \Phi}$ factor ensures that, when evaluated on the trivial topology on the worldsheet, it captures the divergences in the torus. If the dilaton field displays a non-trivial background profile $\Phi(X)$, not only a zero mode $\lambda$, we expect that replacing $\Phi$ with $\Phi(X)$ would lead to a first term in an $\alpha'$ expansion of the term. This term modifies the $\beta$-functional (we will refer from now on to those $\beta$-functionals modified due to the presence of string loop corrections as $\Tilde{\beta}$) associated with the metric through the addition of a term $\delta \beta^G_{\mu \nu}$ to the functional $\beta^{G}_{\mu \nu}$ above
\begin{align}
    \tilde{\beta}^{G}_{\mu \nu} = \beta^G_{\mu \nu} + \delta \beta^G_{\mu \nu},
\end{align}
which looks like a cosmological constant term, i.e. 
\begin{align}
    \delta \beta^G_{\mu \nu} =  C e^{- \Phi} G_{\mu \nu} ,
\end{align}
where $C$ is an arbitrary constant that arises in the renormalization procedure. On equal footing, an additional contribution to the dilaton, which we call $\delta \beta^{\Phi}$ will also appear, although it is hard to evaluate explicitly. Instead, it is easier to obtain it by applying a consistency argument~\cite{Fischler1986,Fischler1986b,Fischler1987}. As we have argued above, in principle the vanishing of the modified $\tilde{\beta}^{\Phi}$-functional through string loop corrections is not independent of the vanishing of the $\tilde{\beta}^{G}_{\mu \nu}$ functional. As we have seen, in the CFT computation, it being constant is precisely a consequence of the vanishing of the remaining $\beta$-functionals. By this consistency condition, it is possible to derive an equation for the $\tilde{\beta}^{\Phi}$-function. 

Taking the divergence of the $\tilde{\beta}^{G}_{\mu \nu}$ and simplifying it through Bianchi identities and using also the vanishing of $\tilde{\beta}^{G}_{\mu \nu}$ itself, we find: 
\begin{align}
    \nabla^{\mu} \tilde{\beta}^{G}_{\mu \nu} =  \nabla_{\nu} \left( \frac{1}{2} R \left( \boldsymbol{G} \right) - \nabla^2 \Phi - \frac{1}{2} \left( \nabla \Phi \right)^2 \right)
\end{align}
This leads us to the following $\tilde{\beta}^{\Phi}$ functional for the dilaton field:
\begin{align}
    \tilde{\beta^{\Phi}}=  \alpha' \left[ -  R \left( \boldsymbol{G} \right) + 2 \nabla^2 \Phi + \frac{1}{2} \left( \nabla \Phi \right)^2 \right] , 
\end{align}
which knowing that is a constant, can be safely chosen to be equal to zero. In case that we were dealing with strings in non-critical dimension, an additional $D-26/6$ factor should be included arising from the $bc$-ghost system contribution to the Weyl-anomaly at the string tree level. Notice that we have introduced $\alpha'$ as a dimensionful parameter. Once we have reached this point, it is better to pause and recapitulate what we have done until now. We began analyzing the $\alpha'$-expansion of the sigma model describing the propagation of strings in arbitrary backgrounds. We determined the $\beta$-functionals of the Weyl anomaly to the lowest order. Then we jumped into the problem of including string-loop corrections that should clearly modify the constraints that the background fields should obey. For the purpose of including such corrections, we noticed that the divergences arising from the string loops can be absorbed into a renormalization of the background fields $G_{\mu \nu}$ and $\Phi$. Hence, up to this point we have found a set of equations that these background fields need to obey for the consistent propagation of the strings. 

\subsection{EFTs for the theory }
\label{Subsec:EFTs}

The consistency equations that we found arising from the Weyl anomaly cancellation and the cancellation of the divergences from string loop corrections resemble a lot the equations of motion of a given field theory for $\mathcal{G}_{\mu} (X)$ and $\Phi(X)$: 
\begin{align}
    & \Tilde{\beta}^{G}_{\mu \nu} = R_{\mu \nu} \left( \boldsymbol{G} \right)- \nabla_{\mu} \nabla_{\nu} \Phi +  C e^{-\Phi} G_{\mu \nu}  + \order{\alpha'}, \\
    & \Tilde{\beta}^{\Phi} = \frac{D-26}{6} + \alpha' \left[ - R \left( \boldsymbol{G} \right) + 2 \nabla^2 \Phi +(\nabla \Phi)^2 \right] +  \order{\alpha^{\prime 2}}.
\end{align}
Setting $C=0$ corresponds to omitting the string loop corrections. The natural question is then whether it is possible to obtain an effective action whose dynamics correctly reproduce these equations. In addition, such effective action needs to correctly account for the scattering amplitudes involving only massless excitations of the string (to the lowest order in the $\alpha'$ expansion) in order to be a sensible action. There are (at least) two effective actions that fullfill these criteria: match the scattering amplitudes involving gravitons and dilatons and their equations of motion give rise to the $\beta$-functionals. These two actions correspond to a GR-like EFT and a UG-like EFT. The GR-like EFT can be given as:
\begin{align}
    S_{\text{eff}}^{\text{GR}} = \frac{1}{2 \kappa^2} \int d^{D+1} X \sqrt{-G} e^{ \Phi} \left(  - \frac{(D-26)}{6 \alpha'}  - 2 C e^{-\Phi} + R \left( \boldsymbol{G} \right) + (\nabla \Phi)^2 \right) + \order{\alpha'}.
    \label{Eq:Eff_Action_GR}
\end{align}
From this action principle it is straightforward to obtain the $\beta$-functionals as 
\begin{align}
    & \tilde{\beta}^{\Phi} = - 2 \kappa^2 \frac{ e^{-\Phi}}{\sqrt{-G}} \frac{\delta S^{\text{GR}}_{\text{eff}} }{\delta \Phi} , \\
    & \tilde{\beta}^{G}_{\mu \nu} = 2 \kappa^2 \frac{e^{- \Phi }}{\sqrt{-G}} \left( \frac{\delta S^{\text{GR}}_{\text{eff}} }{\delta G^{\mu \nu}} + \frac{1}{2} G_{\mu \nu} \frac{\delta S^{\text{GR}}_{\text{eff}}}{\delta \Phi } \right).
\end{align}
Furthermore, it is possible to perform a field redefinition to map this action to the Einstein Frame~\cite{Green1987a}. 

Following~\cite{Carballo-Rubio2022} we know that it is also possible to write down an action principle which reproduces the same equations of motion that Eq.~\eqref{Eq:Eff_Action_GR} displays, with the cosmological constant $C$ entering as an integration constant instead of a coupling constant. To be concrete, we can write down the following action principle: 
\begin{align}
    S_{\text{eff}}^{\text{UG}} = \frac{1}{2 \kappa^2} \int d^{D+1} X \omega e^{ \Phi} \left(  - \frac{(D-26)}{6 \alpha'}   + R ( \boldsymbol{\tilde{G}} ) + ( \tilde{\nabla} \Phi)^2 \right) + \order{\alpha'}. 
    \label{Eq:Eff_Action_UG}
\end{align}
If we compute the variation with respect to $G_{\mu \nu}$ we obtain the traceless version of the equations obtained from Eq.~\eqref{Eq:Eff_Action_GR}. Explicitly, if we define 
\begin{align}
     \frac{\delta S_{\text{eff}}^{\text{UG}}}{\delta G_{\mu \nu}} =  K_{\mu \nu} \left( \boldsymbol{G} \right) - \frac{1}{2}  K \left( \boldsymbol{G} \right) ,
\end{align}
for the variation of $S_{\text{eff}}^{\text{UG}}$ we obtain the following: 
\begin{align}
    \frac{\delta S_{\text{eff}}^{\text{UG}}}{\delta G^{\mu \nu}} = K_{\mu \nu} ( \boldsymbol{\tilde{G}} ) - \frac{1}{D+1} K ( \boldsymbol{\tilde{G}} )  \tilde{G}_{\mu \nu} = 0.
\end{align} 
with $K ( \boldsymbol{\tilde{G}} ) = \tilde{G}^{\mu \nu} K_{\mu \nu} ( \boldsymbol{\tilde{G}} ) $. Upon taking the divergence and using the generalized Bianchi identities for the corresponding tensor $K$ entering the equations (see~\cite{Carballo-Rubio2022} for further details) we find: 
\begin{align}
    E_{\mu \nu} = K_{\mu \nu} ( \boldsymbol{\tilde{G}} ) - \frac{1}{2} K ( \boldsymbol{\tilde{G}} ) \Tilde{G}_{\mu \nu} + C \tilde{G}_{\mu \nu} = 0. 
\end{align}
Again, a suitable combination of these equations with the equation obtained from the equation of motion for $\Phi$ we find: 
\begin{align}
    & \tilde{\beta}^{\Phi} = - 2 \kappa^2 \frac{ e^{-\Phi}}{\omega} \frac{\delta S^{\text{UG}}_{\text{eff}} }{\delta \Phi} , \\
    & \tilde{\beta}^{G}_{\mu \nu} = 2 \kappa^2 \frac{e^{- \Phi }}{\omega} \left( E_{\mu \nu} + \frac{1}{2} \tilde{G}_{\mu \nu} \frac{\delta S_{\text{eff}}}{\delta \Phi } \right),
\end{align}
confirming our claim that the Unimodular Gravity action~\eqref{Eq:Eff_Action_UG} reproduces the $\beta$-functionals. Notice that this effective action does not only reproduce the $\beta$-functionals but it also reproduces all of the scattering amplitudes involving massless excitations of the string (graviton and dilaton asymptotic states), as derived following the procedure sketched in the previous section. In that sense, both actions reproduce the desired properties and hence none of them is preferred over the other one from the perspective of using them as EFTs for the massless modes of the string.

\section{Conclusions}
\label{Sec:Conclusions}

We have analyzed the embedding of UG in string theory from the point of view of the consistent quantization of the strings in an arbitrary background. Furthermore, we have followed the proposal by Susskind and Fischler towards cancelling divergences arising from string loops with suitable counterterms in the non-linear sigma model. Our analysis here does not unveil any preference for UG or GR as a low energy description of string theory. This ties up the loose ends that were not analyzed in~\cite{Carballo-Rubio2022}, regarding the embedding of UG in string theory. To put it explicitly: both UG and GR are equally valid as low energy descriptions of the massless modes of string theory and none of them seems to be preferred over the other one. 

Regarding future directions of work, we recall that our analysis here has focused on bosonic string theory. At first sight, the extension to superstring theory seems straightforward although subtleties may arise in a careful study. Previous considerations of supergravity in a UG-like context suggest that some of the vacua may spontenously break SUSY and hence both theories may develop a potential inequivalence at the quantum level~\cite{Anero2019,Anero2020,Bansal2020}. Although there is no analysis of the global degrees of freedom in such contexts, it should be mentioned that it seems possible that a careful implementation of SUSY in that contexts requires also from a fermionic global degree of freedom, which is the responsible for the apparent SUSY-breaking presented there. 

A second direction of work that is worthwhile exploring is that of non-perturbative definitions of string theory and its interplay with UG. For instance, the gauge/gravity correspondence (also called usually AdS/CFT)~\cite{Maldacena1997,Aharony1999,Polchinski2010} and matrix models~\cite{Ginsparg1993}, among them probably we could highlight the BFSS matrix model~\cite{Banks1996}. In such contexts, we have not explored whether it is easy or not to accomodate a UG principle instead of a GR principle.

String Field Theory (SFT) provides a framework in which it is possible to analyze some features of string theory in a simpler way~\cite{Erbin2021,Sen2019}. For our purpose of discerning whether UG or GR are preferred somehow as the EFTs for the massless string states, it is clear that we have to concentrate on Closed String Field Theory (CSFT) since closed strings are the ones carrying the graviton excitations. There are in principle two ways in which CSFT can shed light on this problem. The first of them is through the study of some non-perturbative formulation of the theory. However, as far as we know, there does not exist any analysis of non-perturbative effects in CSFT that can be used to discern whether UG or GR is preferred over the other one. The second way is the study of background independence of string theory. CSFT has been shown to be background independent in the following sense: although the quantization of the worldsheet requires that we choose a given particular background to which we associate the worldsheet CFT, from the point of view of SFT, all of them are related through a string field redefinition as long as they correspond to marginal deformations~\cite{Sen2017}. However, although this tells us that the different CFTs formulated in different backgrounds are equivalent, it is still possible to write down an effective action for the massless string states that includes a background structure: the background independence only refers to the fact that the worldsheet formulation depends on a background structure. Thus, further work in this direction is required to understand whether a UG principle is preferred over a GR principle.

\begin{acknowledgements}
The authors would like to thank Carlos Barcel\'o and Ra\'ul Carballo-Rubio for collaboration in early stages of this project and invaluable discussions during the preparation of the manuscript. We would also like to thank Tom\'as Ort\'in for helpful conversations. Financial support was provided by the Spanish Government through the projects PID2020-118159GB-C43, PID2020-118159GB-C44, and by the Junta de Andaluc\'{\i}a through the project FQM219. GGM acknowledges financial support from the grant CEX2021-001131-S funded by MCIN/AEI/10.13039/501100011033. GGM is funded by the Spanish Government fellowship FPU20/01684.
\end{acknowledgements}

\newpage

\appendix 

\bibliography{ug_strings_biblio}

\end{document}